# Fabrication of robust capsules by sequential assembly of polyelectrolytes onto charged liposomes

Marta Ruano[1,†], Ana Mateos-Maroto,[1] Francisco Ortega[1,2]*, Hernán Ritacco[3], José E. F. Rubio[4], Eduardo Guzmán[1,2], Ramon G. Rubio[1,2]

[1] Departamento de Química Física, Facultad de Ciencias Químicas, Universidad Complutense de Madrid, Ciudad Universitaria s/n, 28040-Madrid, Spain
[2] Instituto Pluridisciplinar, Universidad Complutense de Madrid, Paseo Juan XXIII 1, 28040-Madrid, Spain
[3] Instituto de Física del Sur (IFISUR)-Universidad Nacional del Sur, Av. Alem 1253, 8000-Bahía Blanca, Argentina.
[4] Centro de Espectroscopía y Correlación, Universidad Complutense de Madrid, Ciudad Universitaria s/n, 28040-Madrid, Spain

* To whom correspondence must be addressed: fortega@ucm.es

† Current address: Strathclyde Institute of Pharmacy and Biomedical Sciences, University of Strathclyde. 27 Taylor Street, G4 0NR-27 Taylor Street. G40NR-Glasgow, United Kingdom

# Abstract

This work presents a simply methodology for coating small unilamellar liposomes bearing different degrees of positive charge with polyelectrolyte multilayers using the sequential Layer-by-Layer (LbL) deposition method. The liposomes were made of mixtures of DOPC (1,2-dioleyl-*sn*-glycero-3-phosphocoline) and DODAB (dimethyl dioctadecyl ammonium bromide), and coated by alternated layers of the sodium salt of poly(4-styrenesulfonate), PSS, and poly(allylamine), PAH, as polyanion and polycations, respectively. The results shows that the zeta potential of the liposomes was not very sensitive to the mole fraction of DODAB in the membrane, $X_D$, in the range $0.3 \leq X_D \leq 0.8$. We were able to coat the liposomes with up to four polymer bilayers. The growth of the capsules size was followed by dynamic light scattering, and in some cases by cryo-TEM, with good agreement between both techniques. The thickness of the layers, measured from the hydrodynamic radius of the coated liposome, depends on the polyelectrolyte used, so that the PSS layers adopt a much more packaged conformation than the PAH layers. An interesting finding is that the PSS amount needed to reach the isoelectric point of the capsules increases linearly with the charge density of the bare liposomes, whereas the amount of PAH does not depend on it. As expected, the preparation of the multilayers has to be done in such a way that when the system is close to the isoelectric point the capsules do not aggregate. For this, we dropped the polyelectrolyte solution quickly, fast stirring and using dilute liposome suspensions. The method is very flexible and not limited to liposomes or polyelectrolyte multilayers, also coatings containing charged nanoparticles can be easily made. Once the liposomes have been coated, lipids can be easily eliminated, giving rise to polyelectrolyte nano-capsules (polyelectrosomes) with potential applications as drug-delivery platforms.

# 1. Introduction

The research on encapsulation and controlled release of active molecules, e.g. drugs, cosmetics, or pesticides, have undergone an important growth in the last two decades [1,2,3,4]. The concept of drug delivery is based on the maximization of drug efficacy, with minimal side effects. This makes the drug safer and more comfortable for patients to use. However, the preparation of drug delivery systems must face, in many cases, a very important problem related to the fact that most of the drugs are hydrophobic and have to be delivered in a water rich environment (the human body). Moreover, in most instances, depending on the type of drug administration, they have to be protected against aggressive conditions e.g. the low pH of the stomach or the adsorption of some molecules in the gut, which can be facilitated by protecting the drugs inside supramolecular aggregates [5]. However, maintaining the appropriate level of drug concentration in the blood stream during a long time requires the use of structures in which the drugs are embedded, that protect them from aggressive pH conditions, and that allow one to use quantities high enough of hydrophobic drugs in a hydrophilic environment.

Liposomes have been extensively used in drug delivery because their membrane is formed by phospholipids as the cell membrane, thus being biocompatible. Furthermore, their structure allow them to be used for including both hydrophobic and hydrophilic drugs, having good biocompatibility and increasing the efficacy and therapeutic index of the drugs, whereas their toxicity is reduced [6]. Indeed, they have been used for the delivery of vaccines, enzymes or vitamins [7]. The main limitation in the use of liposomes is that they are not stable to changes in the temperature or to aggressive environments such as those existing in the stomach, which limits their use for oral administration of drugs [8].A possible method to extend the use of liposomes for drug delivery is to protect them against aggressive environments [9,10,11,12]. Other possibilities are: a) to provide them higher versatility, e.g. decorating their surface with moieties able to recognize specific targets in some cells, e.g. tumoral cells; b) to increase the number of possible drugs and the amount stored in the membrane or c) to tune the delivery rate by coating the liposomes with other motives, such as polyelectrolytes, nanoparticles or smaller liposomes [13].

There are currently several papers dealing with the coating of liposomes using a single layer of polyelectrolyte [14,15,16,17,18,19,20]. There are also quite a few works dealing with the fabrication of LbL decorated liposomes and their applications. However, most of them have overlooked any detailed analysis of the physico-chemical aspects governing the assembly process, which is required for optimizing the potential applications of these systems [21,22]. Furthermore, it has been also reported in the literature the fabrication of LbL films onto other types of soft nanosurfaces [23] or systems in which liposomes are embedded within flat polyelectrolyte multilayers [24,25]. Scheme 1 shows a sketch presenting some examples of LbL films deposited onto macro- and nano-surfaces.

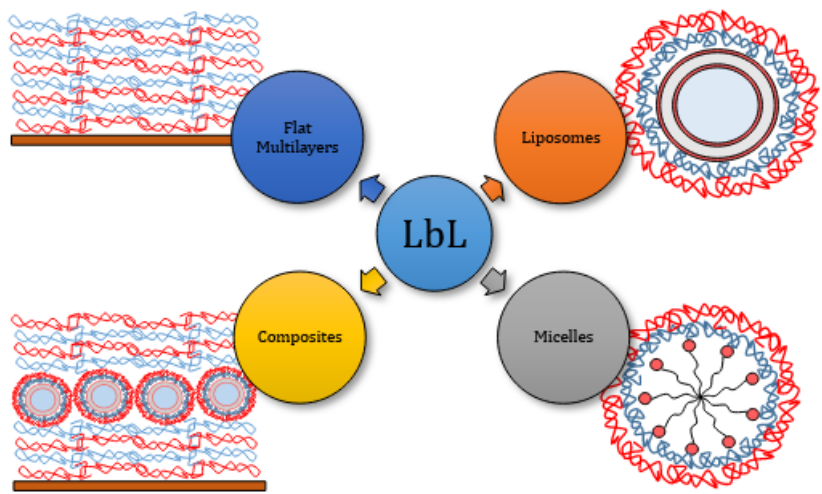

Scheme 1.-Sketch of examples of LbL film deposited onto macro- and nano-surfaces.

A few years ago we briefly described a method for building rather stable nanocapsules by coating liposomes with polyelectrolytes taking advantage of the versatility and modularity of the electrostatic Layer-by-Layer self-assembly (LbL) method [26,27.28]. The groups of Lindman [29,30] and Fukui [31,32] have also shown that the use of the LbL method to coat liposomes with polyelectrolyte multilayers opens new opportunities for fabricating stable drug delivery system, with tunable release profiles. Furthermore, the group of Caruso demonstrated that it is possible to build complex hierarchical multicapsules based on vesicles, capsosomes and liposomes using the LbL method [33,34,35,36]. However, no detailed study of the effect of the charge of the membrane on the coating process has been reported despite that it is well known that it strongly affects the behavior of flat polyelectrolyte multilayers, the mechanical properties of the membranes, and their capacity for storing and delivering drugs [37,38]. Also the growth of the thickness of the coating film as the number of layers increases, the behavior of the zeta potential as subsequent layers are added, and therefore the system stability depends on the template charge density, at least for the first few layers. In the case of flat substrates it was found that increasing the charge density of the substrate leads to an increase of the amount of polyelectrolyte adsorbed [39]. It is worth mentioning that we have demonstrated that the growth of polyelectrolyte multilayers on flat hard substrates and on a fluid/air interface do not present any significant difference [40], therefore it is expected that the above conclusions can be valid for the deposition of LbL films onto a phospholipid bilayer. This can modify the thickness of the multilayers, and therefore the diffusion of molecules through it. In this work we will describe in detail the coating procedure based on the LbL method, and will address the effect of the charge density of the liposome membrane on the coating process. For this purpose, cationic liposomes formed by mixtures of DOPC and DODAB containing different compositional ratios will be studied. The polyelectrolytes PSS [sodium salt of poly(styrene sulfonate)], and PAH [poly(allyl amine)] have been chosen because their behavior in forming polyelectrolyte multilayers on flat surfaces is well documented in the literature [41,42], and therefore a detailed comparison can be done.

It should be stressed that LbL coated liposomes offer advantages in relation to other colloidal systems, such as polymersomes or hollow floating LbL layers. LbL coated liposomes offer up to three different environment to include molecules with different philicity/phobicity: (i) an internal aqueous cavity; (ii) the hydrophobic environment formed by the lipids, and (iii) the external polyelectrolyte shell. Upon the encapsulation of the molecules, these may be released sequentially in a sequential way by combining the breaking of the liposomes and the erosion of the polyelectrolyte. Furthermore, the hierarchical organization of these systems allows combining different functionalities and processes within the assembled systems. Last but not least, the sequential deposition of the polyelectrolyte layers allows tuning the rigidity of the shell almost at will by the choice of the assembled blocks and the number of deposited layer.

## 2. Experimental

### 2.1. Chemicals

1,2-dioleoyl-*sn*-glycero-3-phosphocholine (DOPC) was purchased from Avanti Polar Lipids, Inc. (Alabasted, AL, USA) with a purity higher than of 99%, and stored at $-20°C$. Dimethyldioctadecylammonium bromide (DODAB) was purchased from Sigma-Aldrich (Saint Louis, MO, USA) with a purity higher than 98%, and stored at $25°C$. Poly (sodium 4-styrenesulfonate) (PSS) with a molecular weight of 70 kDa (340 monomers/chain) and polyallylamine (PAH) with a molecular weight of 17 kDa (300 monomers/chain) were supplied by Sigma Aldrich (Saint Louis, MO, USA). All the chemicals were used without further purification. The ionic strength of the solutions was fixed using NaCl with purity higher than 99.99% (Saint Louis, MO, USA). We have used perchloric acid, ascorbic acid and ammonium molibdate from Sigma-Aldrich (Saint Louis, MO, USA) for the determination of the phosphorous content in the liposomes.

All the solutions were prepared by weighting with a precision of $\pm 1$ mg. The water used for cleaning and preparing the solutions was of Milli-Q quality (Milli Q Gradient A10, Millipore Corporation-Burlington, MA, USA), being its resistivity $\Omega > 18$ MΩ · cm and total organic content lower than 6 ppm.

### 2.2. Preparation of the Liposomes

Appropriate amounts of lipids were weighted and dissolved in chloroform (1 mL) to obtain mixtures with the desired composition, i.e. with the desired weight fraction of each individual component. The lipid solutions were homogenized using a vortex, and then the organic solvent was evaporated under a nitrogen stream to produce a dry lipid film, which can be rehydrated with an aqueous solution. During the rehydration process, it is necessary to heat the lipid mixtures above the melting temperature of the lipids used and to homogenize the dispersion by vigorous vortexing. The rehydration process yields a suspension of tiny pieces of membranes and multilamellar vesicles (MLVs). In order to obtain small unilamellar vesicles (SUV), the suspension is subjected to an extrusion process using a Thermobarrel Lipex Extruder from Northern Lipids (Burnaby, Canada) with polycarbonates membranes of 100 nm of diameter. The suspension passed through the membrane several times for ensuring monodisperse liposomes of approximately 100 nm of diameter, during the extrusion process the hydrodynamic radius of the liposomes was checked using dynamic light scattering (DLS) after each five extrusion cycles, in order to optimize the preparation step.

### 2.3. LbL assembly

The liposomes obtained were used as template for the LbL assembly of polyelectrolyte layers. For this purpose, 1 mL of the suspension containing the liposomes (total lipid content 1 mg/mL) is mixed with 1 mL solution of the anionic polyelectrolyte (concentration 1 mg/mL) to form the first layer. Then, the cationic polyelectrolyte was added in excess. This leads to the formation of the second layer of the multilayer and inter-polyelectrolyte complexes (IPECs), formed by the direct interaction between non-adsorbed polyelectrolytes of opposite charge. These complexes precipitate, which enables their separation from the dispersion of coated liposomes by centrifugation at 10000 rpm (3 cycles of 10 minutes). Therefore, even during the preparation process is required the dispersion centrifugation after each deposition cycle for removing the excess of polyelectrolyte as IPECs and to obtain dispersions containing only the prepared capsules, the wasted polyelectrolyte amount is not higher than that wasted using other fabrication processes [28]. The sequential addition of polyelectrolytes with separation of IPECs was repeated several times to fabricate multilayers with the desired number of layers. During the coating steps has been found a lipids lost from 5 to 10% depending on the charge density of the original liposomes. This limits the maximum number of layer to adsorb onto the SUVs templates (around 8-10 polyelectrolyte layers) [26].

### 2.4. Methods

The hydrodynamic radius, $R_H$, of the liposomes coated with polyelectrolyte multilayers was measured by dynamic light scattering, DLS, using an ALV LSE-5003 equipment (ALV Gmbh, Langen, Germany), equipped with an $Ar^+$ laser working at a wavelength of 514.5 nm and a power of 200 mW. The zeta potential, $\zeta$, was calculated from measurements of electrophoretic mobility, $\mu_e$, using the laser Doppler electrophoresis technique (Zeta Nanosizer ZS, Malvern Instruments, Ltd.-Malvern, United Kingdom). The measured $\mu_e$ values were transformed into $\zeta$-potential by the Smoluchowski's relation. The accuracy in the determination of the $\zeta$-potential was better than $\pm 5$ mV. Transmission electron cryo-microscopy images were obtained with a JEOL JEM-1230 microscope (JEOL Ltd., Akishima, Japan). The ESR spectra were obtained with a BRUKER EMX spectrometer (Bruker, Billerica, MA, USA).

The phosphorous titration was done following the method first described by Rouser and later modified by Steward. [43,44] The phosphorous of the phospholipids present in the vesicles is converted to inorganic phosphorous by the addition of perchloric acid, and then a complex with ammonium molibdate and ascorbic acid is formed that can be determined spectroscopically using a UV/visible spectrophotometer (HPUV 8452-Hewlett Packard, Palo Alto, CA, USA), allowing for an estimation of the total phospholipid amount in the liposomes.

## 3. Results and discussion

### 3.1. Phosphorous titration

This is an important step in the present study because we know the initial amount of phospholipids added, but it is necessary to determine whether some phospholipids have been lost during the extrusion process, and also during the coating steps with polyelectrolytes. The latter is important because, as we will discuss, it is necessary to know the amount of polyelectrolyte to add in each step of the multilayer building process. For the sake of example, Figure 1 shows the titration results obtained for liposomes obtained by extrusion of the mixture DOPC:DODAB as a function of the DOPC content (%m DOPC). In the case of mixtures with a weight content of DOPC of 90% the loss can be as significant as 7%. For the mixture the loss of phospholipids decreases steadily as the charge of the liposome increases. Similar results were found for DOPC liposomes. As discussed below, a loss of up to 5% was found after depositing each polyelectrolyte bilayer in the coating process.

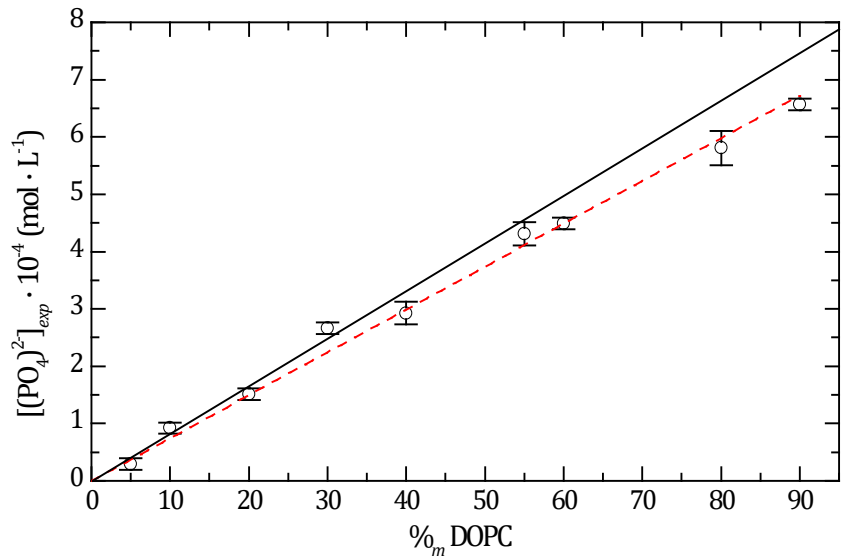

Figure 1.- Phospholipid content in the liposomes after the extrusion process (symbols), evaluated as the phosphate concentration obtained from phosphorus titration, as a function of the initial DOPC content of the mixture of phospholipids. The continuous line represents the expected values if there were no losses, and the dashed line shows the tendency of the experimental data.

### 3.2. Characterization of the bare liposomes

Figure 2b shows the size distribution for the liposomes of DOPC:DODAB 50:50 when they are extruded through membranes of different porous size. For the sake of comparison, it also shows the same information for liposomes of DOPC (Figure 2a). For a given pore size, the mean average hydrodynamic size of the DOPC:DODAB liposomes is slightly shifted towards lower values than those of pure DOPC. This is rationalized considering the higher packing ability of saturated lipids. Thus, the higher the content of DODAB (saturated lipid) the better the packing of the bilayer, and consequently the lower the average size of the liposomes. Furthermore, as expected, the size polydispersity strongly decreases with the porous size.

Figure 3 shows the $\zeta$-potential for DOPC:DODAB liposomes, $R_H = 50$ nm, as a function of DODAB concentration. It is reasonable that the $\zeta$-potential increases sharply as a small amount of DODAB is added, however it remains constant in the interval 10 – 70 wt% of DODAB. This behavior can be explained in terms of the strong condensation of the bromide counterions, that maintains the free charge constant for most of the concentration range. A similar ion condensation effect has already been described in micellar systems [45,46].

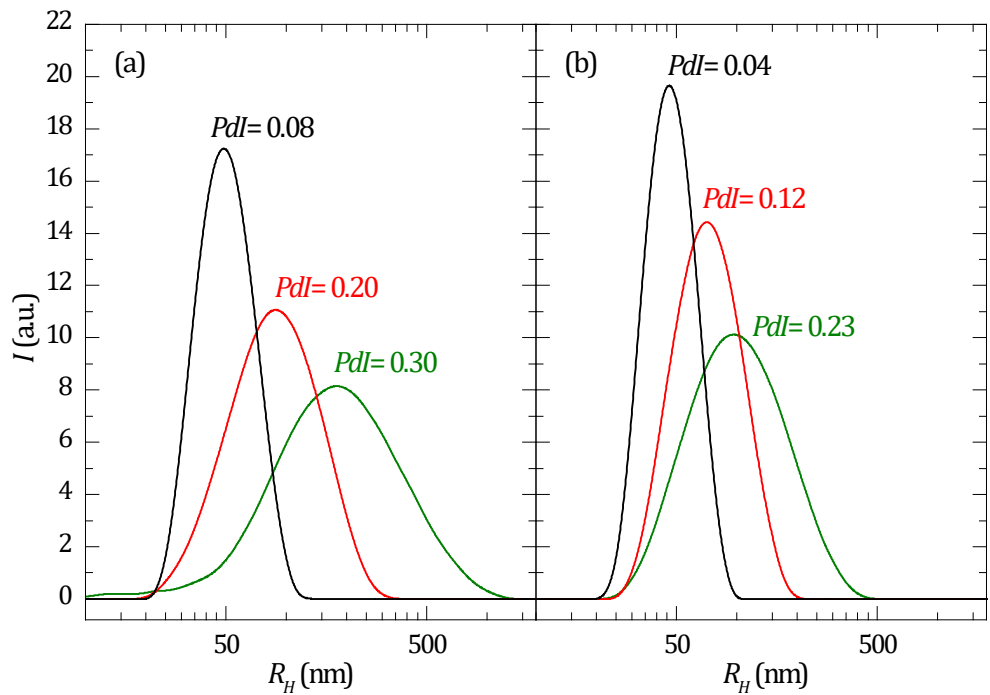

Figure 2.-Hydrodynamic Radius, $R_H$, intensity distributions for liposomes of DOPC (a) and DOPC:DODAB 50:50 (b) obtained by extrusion through filters with different pore diameter: 100 nm (black line), 200 nm (red line) and 400 nm (green line). DLS measurements were performed at [NaCl] = 10 mM, 25 ºC and at a scattering angle of 173º.

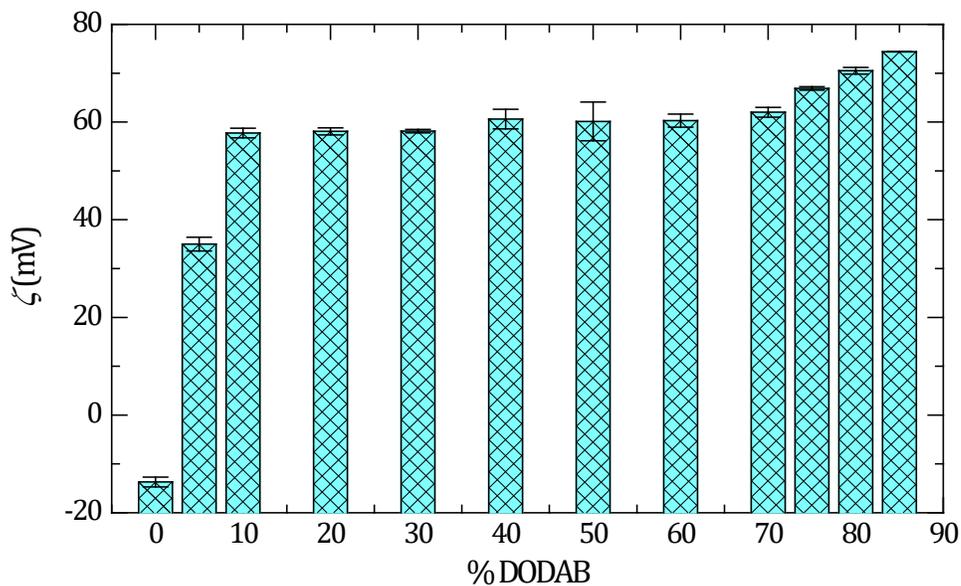

Figure 3.- $\zeta$-potential for DOPC:DODAB liposomes obtained from mixtures with different molar ratios of both lipids. The experiments were performed after the extrusion process, diameter of 100 nm, at 25 ºC and [NaCl] = 10 mM

### 3.3.- Polyelectrolyte coating method

There is no doubt that during the coating of positively charged DOPC:DODAB liposomes with the polyanion PSS the system will pass through the isoelectric point, thus it will become unstable and the liposomes will aggregate. This problem also exists when coating the droplets of oil in water emulsions [47]. To avoid this problem we have worked with dilute suspensions to ensure that the liposomes will be far from each other, and the aggregation process will be slow even when the titration with PSS takes place under stirring. The titration rate and the stirring speed have to be adapted to ensure that the liposomes will have no time to coalesce during the time in which $\zeta$-potential is small. This makes it necessary to check whether the dilution affects the size of the liposomes. We have measured $R_H$ at different liposome concentrations, and found that the size remains constant over the whole concentration range used in this work.

The method followed for coating the liposomes with the first PSS layer is to add a solution of PSS until the isoelectric point has been overcome and charge overcompensation has taken place, at this point the liposome is coated and its surface has negative charge, and an excess of PSS molecules are present in the bulk. The next step is to add a solution of PAH, so that the inter-polyelectrolyte PSS:PAH complexes precipitate, the excess of PAH coats the liposome and overcompensates its surface charge, thus turning it positively charged. The suspension was centrifuged at 10000 r.p.m. during 10 min and the supernatant containing the liposomes is transferred to another beaker. In this process some liposomes are lost trapped by the PSS:PAH complexes as shown by phosphorous titration, in general less than 5% per bilayer. The method can be repeated several times, taking care that the final concentration of liposomes is not too low.

### 3.4.- Characterization of the coated liposomes

Figure 4 shows some of the titration curves obtained for the positively charged liposomes with PSS for different DOPC:DODAB compositions. Near the isoelectric point the values of $\zeta$-potential are hardly reproducible because of the presence of the polyelectrolyte complex. In any case, the inset of Figure 4a shows that the PSS concentration needed for reaching the isoelectric point depends linearly with the DODAB content in the liposome, hence knowing the DODAB composition it is possible to calculate the amount of PSS necessary for neutralization. We found that after charge overcompensation the value of $\zeta$-potential is almost independent of the concentration of each phospholipid. The fact that $C_{PSS}$ ($\zeta = 0$) depends on the charge of the liposome membrane used as template is well known in the construction of polyelectrolyte multilayers on flat surfaces. In this case it is frequently found that the template effect is lost only after 6 or 7 layers [28].

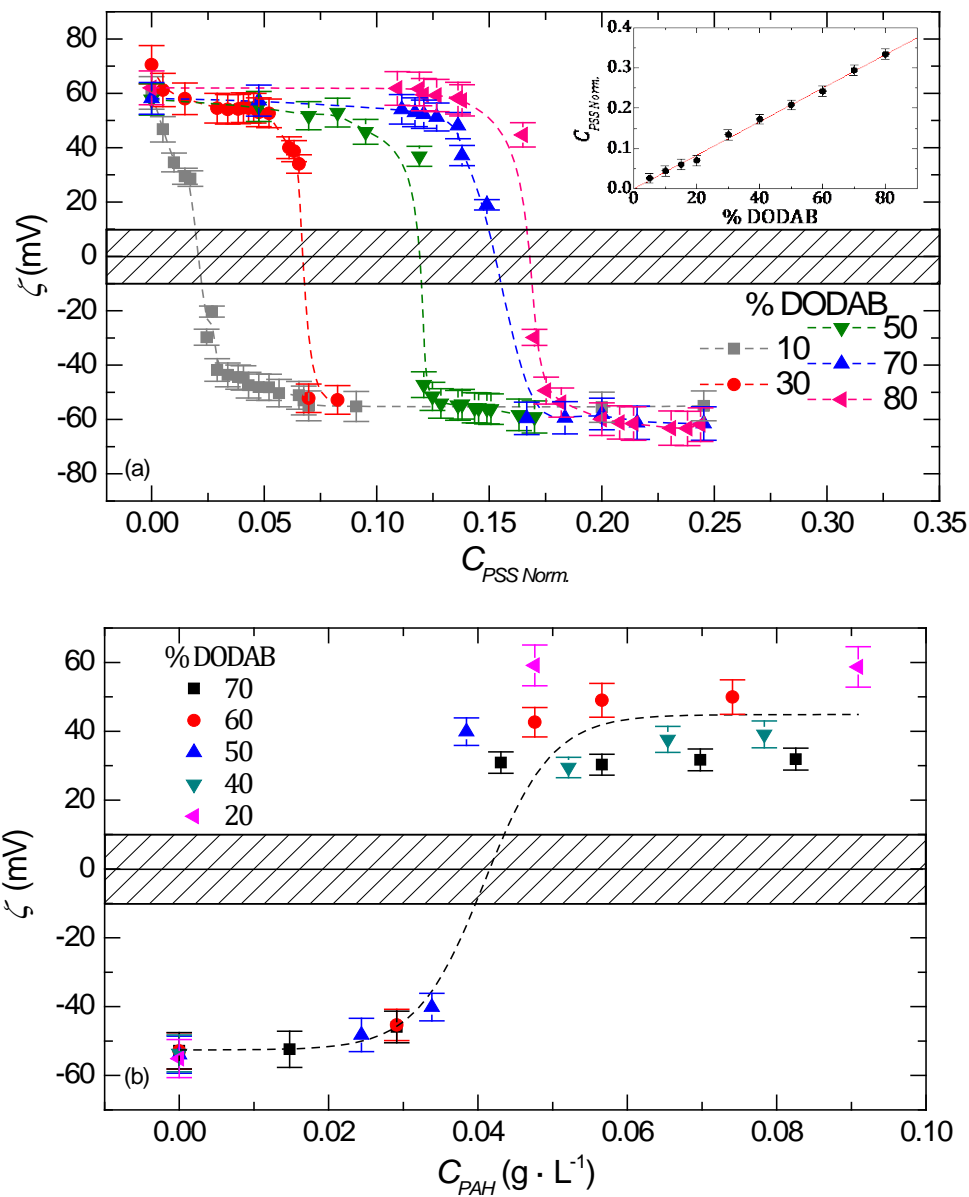

Figure 4.- (a) Dependence of the $\zeta$-potential on the concentration of PSS for vesicles with different concentrations of DODAB. The inserted graph shows the dependence of the normalized amount of PSS needed for reaching the isoelectric point when coating with the first layer of PSS as a function of the percentage of DODAB in the liposome membrane. $C_{PSS\,Norm.}$ is $C_{PSS}$ over the lipid concentration used for preparing the liposomes. (b) Dependence of the $\zeta$-potential on the concentration of PAH for liposomes with different weight content of DODAB. The horizontal bar represents the instability region, where the $\zeta$-potential measurements are not stable. The results correspond to an initial concentration of lipids of 0.5 g·L$^{-1}$. The dash lines are a guide for the eyes.

Figure 4b shows the titration curves for the first layer of PAH. In this case the amount of PAH needed for reaching $\zeta = 0$ does not depend on the DODAB concentration. It is somewhat surprising that no template effect was observed in the second layer of polyelectrolyte. As mentioned above, this is different to what was found for flat surfaces [28]. Even for the second layer of PSS no template effect is observed, thus the amounts of PSS and of PAH necessary for obtaining the isoelectric point are almost independent on the composition of the liposome membrane.

For the sake of example, Figure 5 shows the dependence of the total amount of polyelectrolyte needed for charge overcompensation as a function of the number of layers, $N$, for DOPC:DODAB liposomes with 70:30 compositional ratio (qualitatively similar results were found using DOPC:DODAB liposomes with other compositional ratio, for the sake of example in the Supporting Information, Figure S.1, are included the data for liposomes with 30:70 compositional ratio). The alternating effect with the deposition of polyanion and polycation layers is clearly observed within the multilayer. Whereas the template effect is important for PSS, it is small for PAH, which can be due to the different weight of the electrostatic interactions in the adsorption of both polyelectrolytes. For PSS, a strong polyelectrolyte, the adsorption is mainly driven by electrostatic interactions and it is expected that the charge density of the template plays an important role on the charge overcompensation. However, in the case of PAH, a weak electrolyte, the entropic and specific interactions are important, thus the template effect is reduced [48].

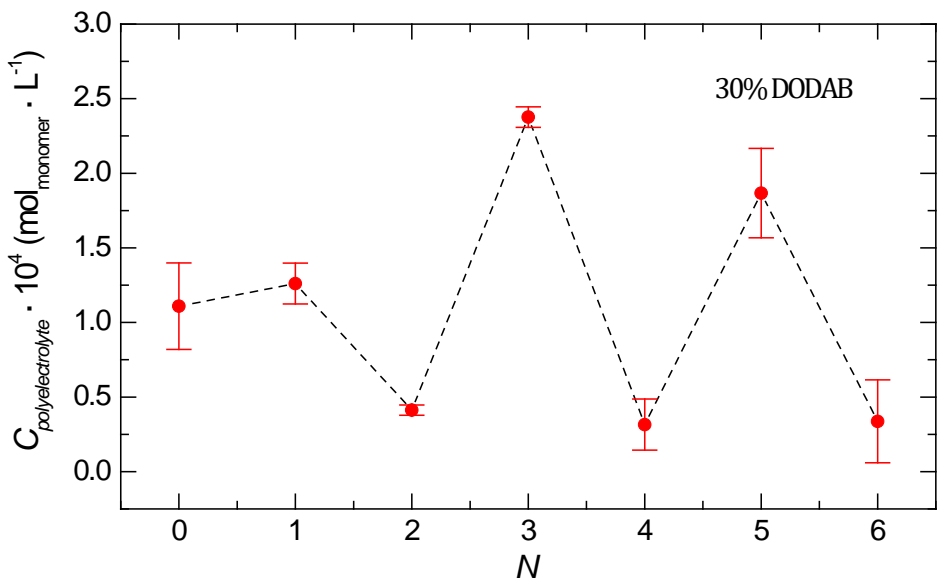

Figure 5.- Amount of polyelectrolyte needed for reaching overcompensation as a function of the number of PSS/PAH layers deposited onto DOPC:DODAB liposomes with 70:30 compositional ratio.

It has been possible to obtain information on the type of charge compensation mechanism, intrinsic or extrinsic, responsible for the formation of the multilayers from the titration curves [49]. First it is necessary to evaluate the concentration of liposomes, $C_{lip}$, and the number of charges at their outer layer, $N_{co}$. For the calculation of $C_{lip}$ we have calculated the number of lipids per liposome, $N_{ph}$, using two different methods. The first method implies the calculation of the total area of the liposome membrane (calculated from the hydrodynamic radius, $R_H$), the average area occupied by a lipid at each DOPC:DODAB composition, and the thickness of the liposome membrane (we have taken $h\sim 5$ nm as in the case of the 1,2-dipalmitoyl-*sn*-glycero-3-phosphocoline (DPPC) bilayer [50], although the final results are not significantly affected in the range 4-7 nm). The second method uses the total volume of the liposome membrane and the average density of the lipid mixture. We have found that both methods agreed within 2%. An illustrative example of $N_{ph} = 128714$ for the calculation using the volume method and 101042 using the method based on the area were obtained for the DOPC:DODAB liposomes with 70:30 compositional ratio. This difference is not relevant for the final calculations. The total amount of DOPC in solution, $C_{DOPC}$, was obtained from the phosphorous titration (see Figure 1), and the DODAB concentration was calculated from $C_{DOPC}$ and the relative concentration of both phospholipids. All the above information, together with $N_{ph}$ allowed us to calculate $C_{lip}$. $N_{ph}$ and $h$ allow one to obtain the number of phospholipids in the outer layer of the membrane. Assuming that there is no preferential distribution of DODAB or DOPC in the inner and outer layers of the membrane, $N_{co}$ was calculated, which together with $C_{lip}$ allowed us to obtain the total number of charges at the outer layer of the liposomes in the whole suspension.

The concentration of PSS necessary for coating the liposome, $C_{PSS}$, (Figure 4a, $\zeta \approx -60$ mV), together with $C_{co}$, makes it easy to calculate the number of monomers necessary for completing the first polymer layer, $N_{PSS}$. The results point out that full charge overcompensation is reached when $N_{PSS}/N_{co} \approx 2.75$, corresponding to an extrinsic compensation mechanism. In addition to the fact that counterions can compensate some of liposome surface charges, one has to consider that the distance of the sulfonate groups in PSS might not fit the average distance between DODAB heads at the outer layer of the membrane, so some polymer charges would not compensate DODAB charges. The ratio $N_{PSS}/N_{co}$ is analogous to the effective $L/D$ ratio usually discussed in the study of the interactions of DNA and vesicles in gene transfection [51].

In a similar way we have calculated the number of PAH monomers, $N_{PAH}$, necessary for overcompensating the charges of the PSS layer. The values $N_{PAH}/N_{PSS}$ plotted in Figure 6a show that the mechanism changes from extrinsic to highly intrinsic as the DODAB content increases.

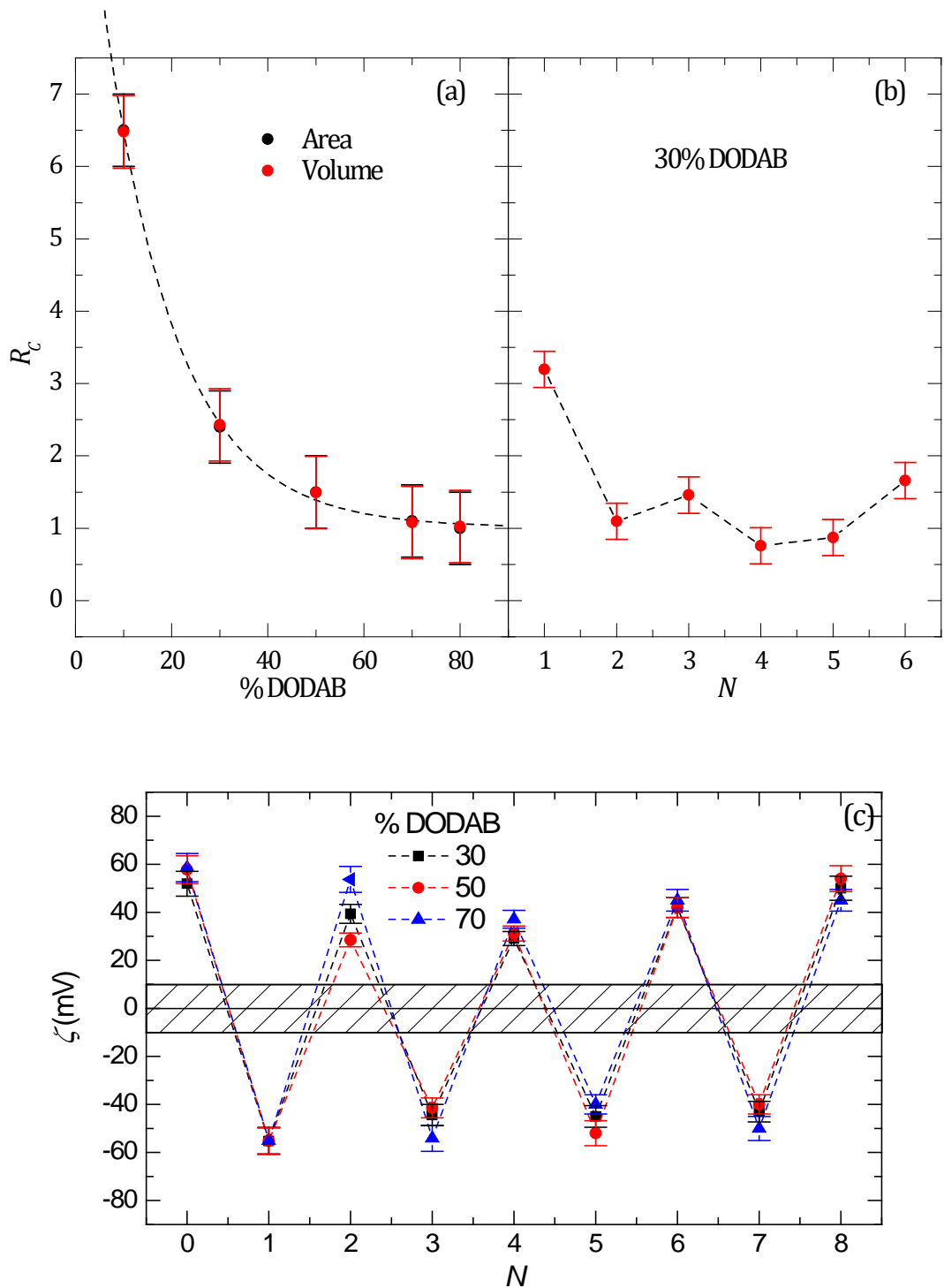

Figure 6.- (a) Ratio between the number of deposited PSS monomers and charges on the external leaflet of the liposome, $R_c$, for full charge overcompensation as function of the weight fraction of DODAB in the liposomes. The labels "area" and "volume" refer to the two ways to calculate the number of charges in the outer layer of the liposome, and the results coincide. (b) Effect of the number of layers on the overcompensation ratio, $R_c$, of polyelectrolyte monomers on each layer for the deposition of a PSS/PAH onto DOPC:DODAB liposomes with 70:30 compositional ratio. (c) Dependence of the zeta potential on the number of PSS/PAH layers adsorbed on the liposomes with different compositional ratio.

Lourenço et al. [52] found that for flat PAH/PSS multilayers on solid substrates the charge compensation mechanism was intrinsic. However, the study of Riegler and Essler [53] on PAH/PSS multilayers adsorbed to a DODAB monolayer at the air/water interface concluded that the compensation mechanism was extrinsic above certain threshold value of the polyelectrolyte charge density, which is fulfilled by PAH under our experimental conditions. However, no one has reported yet the effect of the charge density of the template on the compensation mechanism. These results are unexpected because for some floating and solid supported multilayers both the dependence of the multilayer thickness on the number of polymer layers, and the mechanical properties of the multilayers were the same. The difference between the results reported by Lourenço et al. [52] and by Riegler and Essler [53] might be due to the presence of the DODAB monolayer as template, which would be very important for the coating of liposomes. The present results confirm this template effect. After the first bilayer no effect of the liposome charge density has been observed neither on $C_{PSS}$ nor on $C_{PAH}$. One has to be very careful in extrapolating the above results to other similar systems because it is well known from the results for flat polyelectrolyte multilayers that the compensation mechanism strongly depends on the nature of the polymers, and on other variables such as pH or ionic strength [37,42,49].

The effect of $N$ on the overcompensation ratio is shown in Figure 6b for DOPC:DODAB with 70:30 compositional ratio where it is observed that, except for the first PSS layer, the multilayer presents a mainly intrinsic compensation mechanism. The results clearly show that despite the strong change from intrinsic to extrinsic compensation shown in Figure 6a as the DOPC ratio increases, beyond the second layer the compensation mechanism becomes intrinsic. This behaviour is just the opposite than the one previously reported for multilayers adsorbed onto planar substrates in which increasing the number of layers leads to a change from intrinsic to extrinsic compensation [42]. It should be noted that the behavior for the assembly of multilayers on liposomes with other compositional ratio showed qualitatively similar dependences (see Figure S.2 in Supporting information for the case of DOPC:DODAB liposomes with 30:70 compositional ratio).

### 3.5. Zeta potential and hydrodynamic radius of the capsules

Figure 6c shows the typical oscillatory behavior of the zeta potential as a function of the number of polyelectrolyte layers, $N$, [¡**Error! Marcador no definido.**] for the deposition of LbL multilayers onto liposomes with selected values of charge density, i.e. liposomes containing selected amounts of DODAB (Note: the deposition of LbL layers on liposomes with other charge densities present similar dependences). The results do not evidence systematic effects neither on the charge density of the template nor on the number of polyelectrolyte layers.

The correlation functions obtained by DLS showed an exponential decay both for the bare liposomes, and for the coated ones over a broad range of scattering angles. The single exponential shape is consistent with quite monodisperse samples. From the plot of the inverse of the characteristic decay time *vs.* the square of the wavevector we concluded that the dynamics of the capsules is diffusive in all the cases. The slopes of those linear plots allowed us to calculate the diffusion coefficient, $D$, from which $R_H$ was calculated using the Stokes-Einstein equation. The same conclusion are valid for liposomes with other DOPC:DODAB compositional ratio. It is worth to compare the behavior of $\zeta$-potential and $R_H$ (Figure 7a and b, respectively) because it clearly shows that a maximum in $R_H$ appears at a value of $C_{PSS}$ close to that of the isoelectric point, this maximum in $R_H$ corresponds to instability region, as is expected from the null value of the effective charge of the complexes, i.e. the $\zeta$-potential. Considering that the association process occurs under equilibrium conditions, it would be expected an aggregation reversible process, and the disappearance of the aggregates as result of the charge overcompensation. However, the mixtures of the liposome dispersion and the polyelectrolyte solutions in presence of concentration gradients may result in the formation of kinetically-trapped aggregates similar to that appearing in polyelectrolyte-surfactant mixtures [54,[55],[56]].

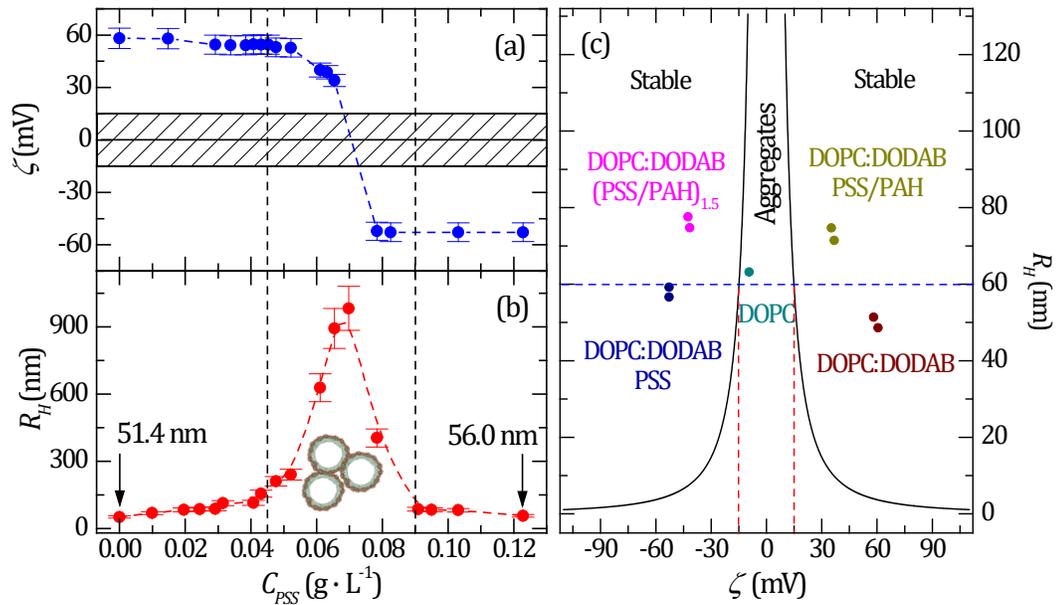

Figure 7.- Dependence of the $\zeta$-potential (a) and the hydrodynamic radius (b) for a suspension of liposomes DOPC:DODAB with 70:30 compositional ratio as a function of the PSS added. Aggregation takes place between the vertical dashed lines. The dashed rectangle on the panel a represents the instability region where the zeta potential measurements are not reproducible. c) $\zeta$-potential dependence of the hydrodynamic radius for all the liposome suspensions studied. The Velegol-Thwar theory predicts that the systems are unstable in the region between the two continuous lines. The horizontal dashed line represents the so-called ideal stability region.

For explaining the aggregation process in the system it has been necessary to use the Velegol-Thwar theory that relates the radius of the aggregates with zeta potential by [44,57]

$$R_H \approx \frac{10 k_B T}{\pi \varepsilon} \left\{ (\zeta^2 + \sigma^2) \ln \left[ 1 - \left( \frac{\zeta^2}{\zeta^2 + \sigma^2} \right) \right] + \zeta^2 \ln \left[ \frac{2\zeta^2 + \sigma^2}{\sigma^2} \right] \right\}^{-1} \qquad (1)$$

where $\varepsilon$ is the dielectric permittivity, and $\sigma$ is the standard deviation of the potential at the surface of the particles, a measure of the heterogeneity of the interaction potential at different points of the surfaces of the liposomes, i.e. heterogeneity of the surface charge density. $\sigma$ is a very important parameter in this theory because it leads to an attractive term in the interaction potential even for particles with charges of the same sign. Figure 8c shows the plot of $R_H$ vs. $\zeta$ for all the DOPC:DODAB capsules studied and the instability region predicted by the theory. It is clear that all the samples are in the stable region. The so-called ideal stability line predicted by the model (continuous lines in the figure) coincides with the capsules coated with one bilayer. It is reasonable that the values for capsules with more layers lie above that line, because the radius increases with the number of layers whereas we have shown that the zeta potential takes only slightly different values (see Figure 6c). The increase of the thickness of the coating takes the system further away from the instability region, which is one of the reasons for coating the liposomes.

For the sake of example, Figure 8a shows the dependence of $R_H$ for all the capsules as a function of $N$ for DOPC:DODAB liposomes with 70:30 compositional ratio. Figure 8b shows the diffusion coefficient of the capsules for different DODAB concentrations. It is clear that except for the bare liposomes, whose values of $R_H$ decrease for higher DODAB concentrations (i.e. the diffusion coefficient increases as show Figure 8b), the values of $R_H$ of the polyelectrolyte-decorated liposomes appears rather independent of the lipid composition of the bilayer. The effect of the charge density on the size of the bare liposomes can be understood in terms of the membrane rigidity associated with the incorporation of DODAB. Thus, lipid bilayers with a high content on DODAB present higher rigidity due to the saturated hydrophobic chains of this lipid that favors the compaction of the molecules within the leaflet. However, as the DODAB content decreases, the lipid bilayer becomes more flexible, and hence the liposome can be deformed instead of broken and re-assembled during the extrusion process, which leads to the formation of bigger liposomes. The presence of DOPC makes the membranes more flexible, and consequently the polydispersity of the liposomes becomes higher after the extrusion process, leading to higher average $R_H$ values, as can be observed from the hydrodynamic Radius, $R_H$, intensity distributions displayed in Figure 2. Furthermore, the polyelectrolyte amount required for the fabrication of the first layer onto the nude liposome also increases with the charge density of the bilayer, and hence it is possible to assume that the adsorption of the first layer appears dependent on the DODAB content. However, the $R_H$ of the liposomes decorated only with a PSS layer is the same independently on the charge density of the bare liposome used as substrate. Thus, the thickness of the first PSS layers appears dependent on the charge of the liposome, the higher the charge of the liposome the higher the thickness of the first PSS layer. This is clear from the inset in Figure 8, where for the sake of example the dependence on the thickness of the first PSS layer on the DODAB fraction in the membrane is displayed for some selected PSS-decorated liposomes. It should be noted that once the first PSS layer is deposited the growth of the multilayer become rather independent on the characteristics of the bare liposome. It is also observed that the increase of the radius of the capsule after the deposition of a PAH layer is almost fourfold that of PSS. In the case of PAH/PSS multilayers grown on a DODAB monolayer at the air/water interface, Riegler and Essler [53] found that the relative increase of the thickness after adding a PAH or a PSS layer depends strongly on the ionic strength, and the results obtained in this work agree reasonably well with such a picture. However, Guzmán et al. [58] reported smaller differences for the same multilayer on a solid substrate over a broad range of ionic strengths.

A representative example of cryo-TEM images of the capsules are shown in Figures 8c and d for DOPC:DODAB liposomes with 40:60 compositional ratio. The diameter of the bare vesicle is $95 \pm 6$ nm in good agreement with the results obtained by DLS. From the images we have estimated the thickness of the membrane for both the bare liposome of for the one coated with four layers, and we obtained $14 \pm 4$ nm per bilayer. Although the agreement with the results shown in Figure 9a is reasonable, the estimation from cryo-TEM has to be taken only as semi-quantitative.

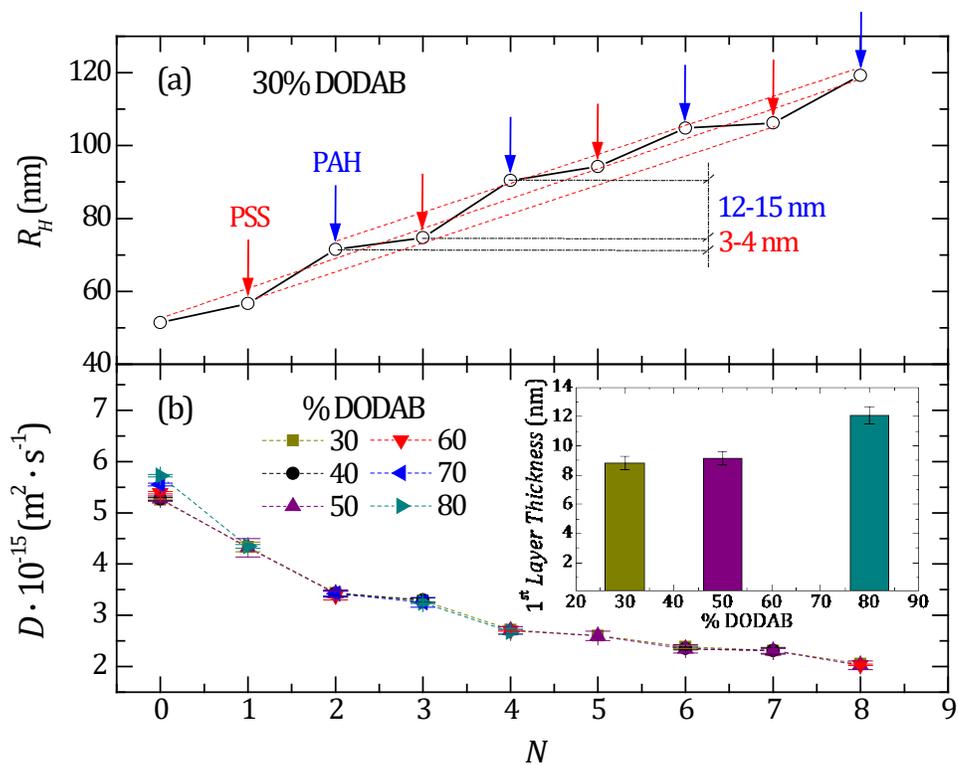

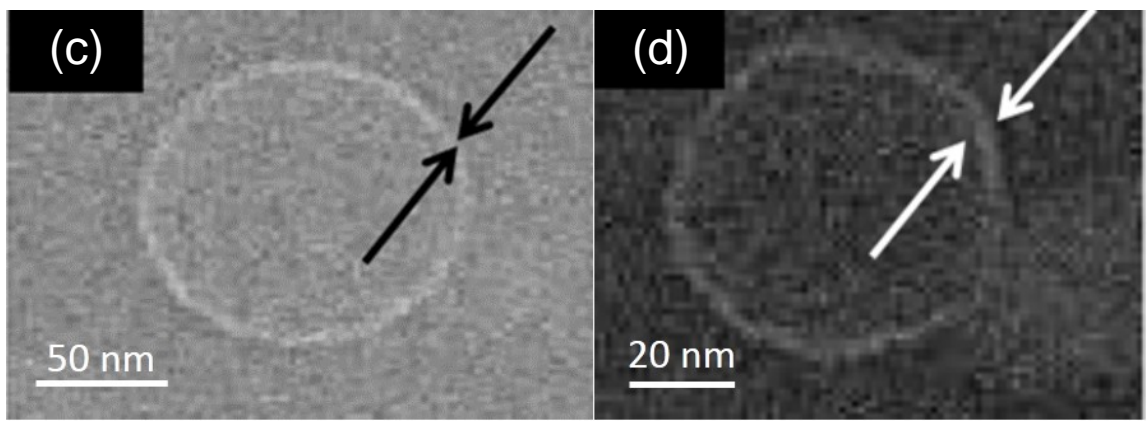

Figure 9.- (a) Dependence of the hydrodynamic radius of the coated liposome as a function of the number of layers. $N = 0$ corresponds to the bare liposome. The DOPC:DODAB compositional ratio is 70:30. (b) Variation of the Diffusion Coefficient, $D$, with $N$. In the bare liposome $D$ changes with the % of DODAB. The inset shows the dependence of the thickness of the first PSS layer on the DODAB content of the bare liposome for some selected percentages of DODAB. on the template liposome. (c) Cryo-TEM image of bare liposome. (d) Cryo-TEM image of liposome coated with four layers of PSS/PAH.

### 3.6. Stability of the capsules

An important issue for the possible use of the capsules is their stability. The results of DLS and $\zeta$-potential in Figure S.3 (see Supporting Information) show that the samples are stable after five weeks, both coated by a single layer of PSS or by four bilayers. However, after seven weeks the $\zeta$-potential of the liposomes coated with a single PSS layer starts to change and becomes less negative, indicating that the system is starting to move towards the instability region (see Figure 9c). This was confirmed by the appearance of aggregates that shift the average value of $R_H$ obtained by DLS and the width of its distribution towards higher values. We found that when the outermost layer was PAH the capsules were slightly less stable than when it was PSS, a similar effect was reported by Cuomo et al. [29,30]. It is important to remark that the increase of stability induced by the coating does not only refer to preventing the aggregation, but also to the chemical stability. Bare liposomes undergo easily oxidization after a couple of week. However, the deposition of polyelectrolyte LbL layers on liposome surfaces prevents the oxidation, with the stability being extended up to two months.

We have remarked that the coating procedure is effective only if the liposome suspension is diluted, otherwise the liposomes aggregate when $\zeta \approx 0$. Coated liposomes allow one to concentrate the suspension and simultaneously to eliminate from the solvent the last polyelectrolyte added by ultrafiltration. Depending on the molecular weight of the polymers, two protocols can be used for removing the last polyelectrolyte from the solvent. The first one is dialysis of the concentrated suspension if the radius of gyration of the polymer is significantly smaller than that of the liposome. The second method is to dilute the concentrated suspension using pure water, then to filter it again and repeat the process until the polymer concentration is low enough. Since some capsules can be lost during filtration, the dialysis procedure is preferable. This possibility of concentrating the capsule suspension is very important for drug delivery purposes. It is important to remark that after reduction of the initial volume of the suspension by four times, the values of $R_H$ and $\zeta$-potential were the same than the initial ones. Of course, in these measurements the suspensions were still transparent for allowing one to perform the DLS experiments.

### 3.7. Effect of the coating on the liposome membrane (Electron Spin Resonance, ESR)

It seems reasonable to think that the strong interaction between the first deposited polyelectrolyte layer and the charged heads of the DODAB molecules can affect to the fluidity of the external leaflet of the membrane, and as matter of fact to its dynamics. This means that the deposition of the first polyanion layer, i.e. PSS layer, creates an electrical field that can interact with the DODAB molecules, limiting the mobility of the molecules at the outer leaflet of the membrane. Thus, considering that DODAB tends to rigidify the membrane, it is expected that the adsorption of PSS may modify the microviscosity of the membrane [59]. ESR technique is very sensitive to the structure of the membrane [60], allowing one to measure the relaxation time, $\tau$, that characterizes the motion of a probe containing free radicals. In the present study we have used N-tempoyl palmitamide as probe. Figure S.4 (see Supporting Information) shows a typical set of ESR spectra obtained at two different temperatures (25ºC and 50ºC) for DOPC:DODAB liposomes with 70:30 compositional ratio coated with different number of polyelectrolyte layers. It must be remarked that the height of the spectra is not normalized and correspond to different concentrations of liposomes. However there is a displacement of the field at which the peaks appear depending on whether the polymer layer is PSS or PAH until $N = 4$, afterwards the signal-to-noise ratio is too low to be significant. The so-called melting transition of a DOPC bilayer is -20 ºC whereas that of a DODAB one is 45 ºC, thus experiments were done at 25 and 50 ºC. From the data obtained in each spectra, the relaxation times can be calculated according to the methods proposed by Cruz et al. [61] and Man et al. [62]. Figure 9 shows the relaxation time values obtained as the number of polyelectrolyte layers increase for DOPC:DODAB liposomes with 70:30 compositional ratio, a similar qualitative trend was obtained for liposomes with 30:70 compositional ratio. A first obvious result is that $\tau$ decreases as result of the decrease of viscosity with temperature, i.e. the relaxation times at 50ºC are 2-3 folds lower than those obtained at 25ºC. A second observation is that there is an odd-even effect in which the addition of a layer of PSS makes the lipid membrane less fluid, this effect is stronger below the transition temperature (25ºC). This behavior is compatible with the strong interaction between DODAB molecules in the outer layer of the lipid membrane and PSS monomers when is added (electrostatic interaction) leading to a more rigid environment for the probe, and an opposite effect when a PAH layer is added.

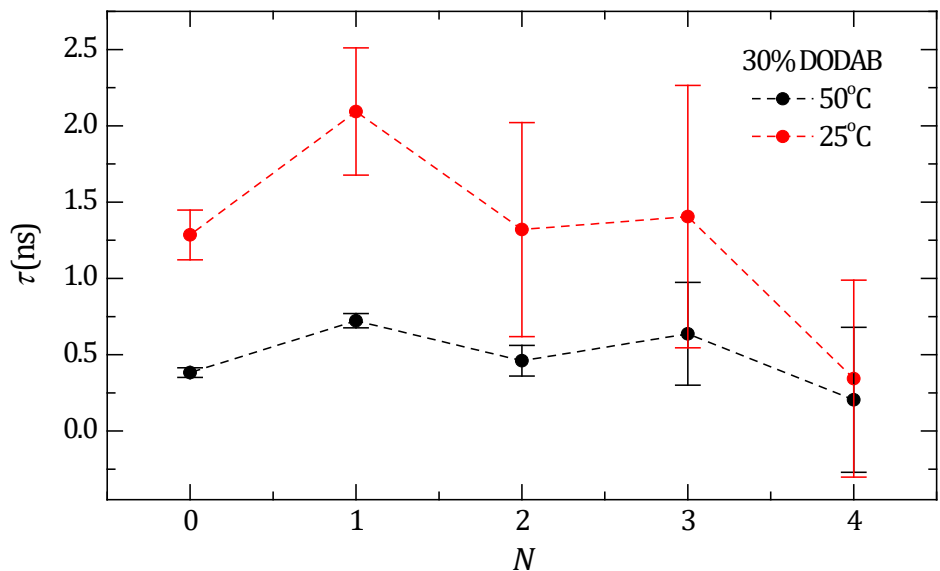

Figure 9.- Effect of the deposition of polyelectrolyte layers onto DOPC:DODAB liposomes with 70:30 compositional ratio on the relaxation times, $\tau$, of the radical probe inside the lipid membrane at to different temperatures.

## 4.    Conclusions

We have described a method based on the electrostatic Layer-by-Layer self-assembly for coating liposomes with LbL polyelectrolyte multilayer. The combination of a phosphorous titration method with DLS data have allowed to conclude that the charge compensation mechanism during the assembly of polyelectrolyte layers can be either intrinsic or extrinsic, depending on the nature of the last deposited polyelectrolyte. We have found that the effect of the surface charge density of the bare liposomes on their size is related to the flexibility of their membranes during the extrusion process. However, after coating with the first polymer layer the charge density of the liposome membrane has no effect on the growth of the subsequent layer or the properties of the final supramolecular system. The possibility of tuning the charge of the membrane at will, whereas the deposition of the layers remains almost unaffected by the charge of the membrane, makes possible to dissolve molecules with different characteristic within the membrane, which may be released without taking care on the conditions requires for tuning the polyelectrolyte shell permeability. Finally, the coating layers increase both the stability against aggregation, but also against oxidation of the lipids. It is true that this study has been limited to the deposition of only four bilayers onto the liposomes. However, it may expected that this number of layer may be enough for enhancing the stability of the liposomes, avoiding their oxidation, and making these systems useful for encapsulation and control delivery of active molecules, taking advantage of the different environments available for including different types of molecules.

Even though a drawback of the method is that one has to start from dilute suspensions of liposomes, it was possible to concentrate the coated liposome suspensions by ultrafiltration. The liposomes in the concentrated suspensions showed the same size and $\zeta$-potential than in the diluted ones.

The use of liposomes as template for creating polymeric capsules instead of solid colloidal particles allows combining the power of liposomes as encapsulation platform with the protection provided by the polyelectrolyte multilayers, which minimizes the destabilization processes of the liposomes dispersion. Furthermore, the use of liposomes as templates allows one to obtain floating polyelectrolyte hollow capsules using mild conditions (dissolution by non-ionic surfactants with reduced toxicity, followed by dyalisis) instead hard physico-chemical treatment, e.g. dissolution in organic solvent or acid solutions.

**Supporting Information**

The Supporting Information is available free of charge at

- Amount of polyelectrolyte needed for reaching overcompensation as a function of the number layers deposited onto DOPC:DODAB liposomes with 30:70 compositional ratio.
- Dependence of the number of deposited PSS monomers and charges on the external leaflet of the liposome for full charge overcompensation on the weight fraction of DODAB in the liposomes, and

- effect of the number of layers on the overcompensation ratio, Rc, of polyelectrolyte monomers for the deposition onto DOPC:DODAB liposomes with 30:70 compositional ratio.
- Long-term stability test for nude liposomes and coated liposomes in terms of the change of the hydrodynamic radius and zeta potential.
- EPR spectra.

**Acknowledgements**

This work in Madrid was funded in part by MINECO and MICINN (Spain) under grants CTQ2016-78895-R and PID2019-106557GB-C21, respectively, by Banco Santander-Universidad Complutense grant PR87/19-22513 (Spain) and by E.U. on the framework of the European Innovative Training Network-Marie Sklodowska-Curie Action NanoPaint (grant agreement 955612). The work of H.R. in Argentina was partially supported by grants PGI-UNS 24/F067 of Universidad Nacional del Sur, PICT-2016-0787 of Agencia Nacional de Promoción Científica y Tecnológica (ANPCyT, Argentina) and by PIP-GI 2014 Nro 11220130100668CO of Consejo Nacional de Investigaciones Científicas y Técnicas (CONICET, Argentina). M.R is grateful to MICYT for a Ph.D. grant. We are grateful to the Centro the Espectroscopía y Correlación of Complutense University of Madrid and Centro de Nacional de Microscopía Electrónica for the use of some of its facilities. Authors thank to Patricia Guisado Barrado (www.patriciaguisado.com) for designing the artwork used to participate in the program for featuring the work as Supplementary Cover in Langmuir.

# Supplementary material
# for
# Fabrication of robust capsules by sequential assembly of polyelectrolytes onto charged liposomes

Marta Ruano[1,†], Ana Mateos-Maroto,[1] Francisco Ortega[1,2]*, Hernán Ritacco[3], José E. F. Rubio[4], Eduardo Guzmán[1,2], Ramon G. Rubio[1,2]

[1] Departamento de Química Física, Facultad de Ciencias Químicas, Universidad Complutense de Madrid, Ciudad Universitaria s/n, 28040-Madrid, Spain
[2] Instituto Pluridisciplinar, Universidad Complutense de Madrid, Paseo Juan XXIII 1, 28040-Madrid, Spain
[3] Instituto de Física del Sur (IFISUR)-Universidad Nacional del Sur, Av. Alem 1253, 8000-Bahía Blanca, Argentina.
[4] Centro de Espectroscopia y Correlación, Universidad Complutense de Madrid, Ciudad Universitaria s/n, 28040-Madrid, Spain

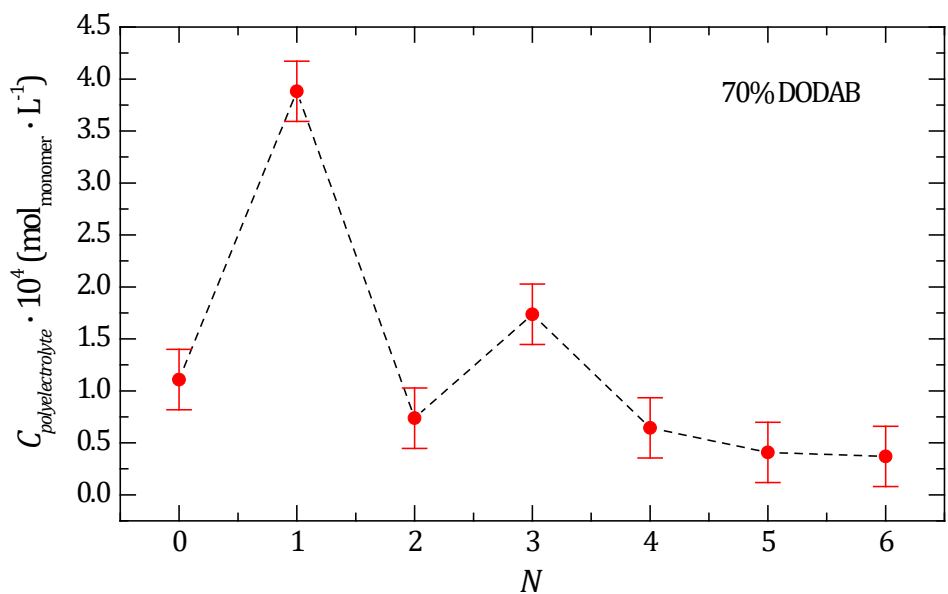

Figure S.1.- Amount of polyelectrolyte needed for reaching overcompensation as a function of the number of PSS/PAH layers deposited onto DOPC:DODAB liposomes with 30:70 compositional ratio.

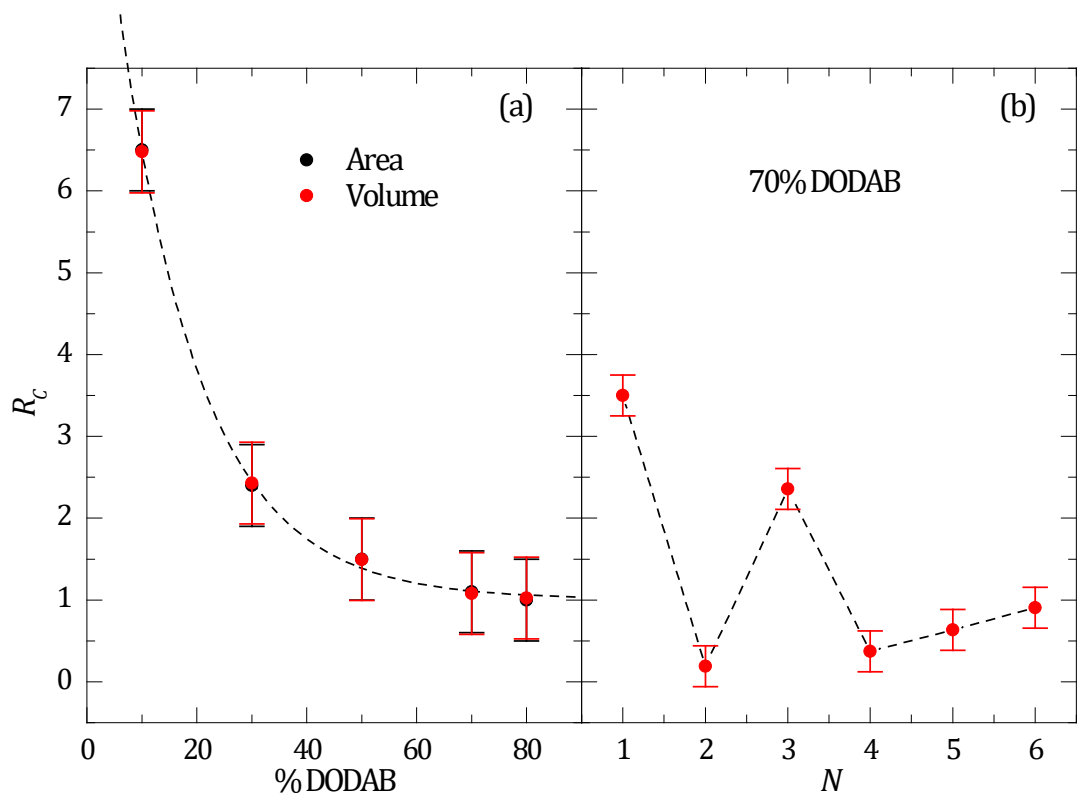

Figure S.2.- (a) Ratio between the number of deposited PSS monomers and charges on the external leaflet of the liposome, $R_c$, for full charge overcompensation as function of the weight fraction of DODAB in the liposomes. The labels "area" and "volume" refer to the two ways to calculate the number of charges in the outer layer of the liposome, and the results coincide. (b) Effect of the number of layers on the overcompensation ratio, $R_c$, of polyelectrolyte monomers on each layer for the deposition of a PSS/PAH onto DOPC:DODAB liposomes with 30:70 compositional ratio.

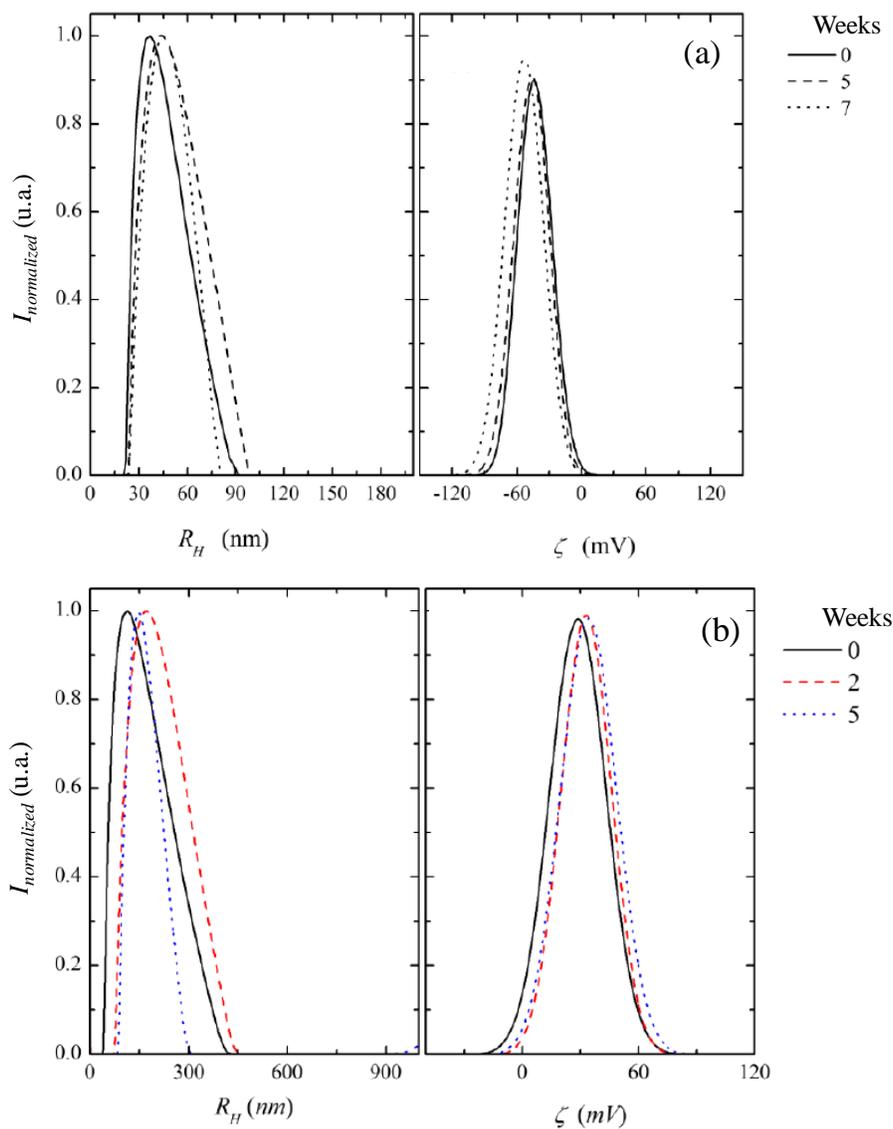

Figure S.3.- Variation of the normalized intensity as a function of the hydrodynamic radius and zeta potential, measured in different times, of (a) liposomes formed by a composition of 40% DODAB and coated with one layer of PSS and (b) coated by two bilayers of PSS/PAH.

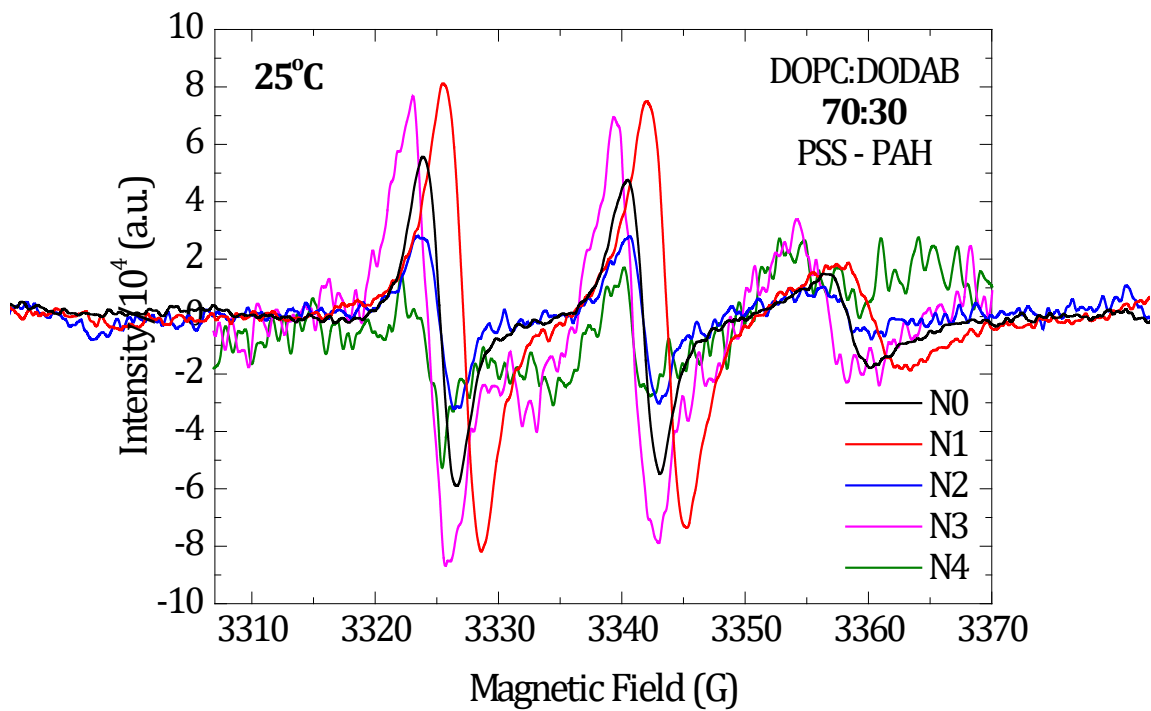

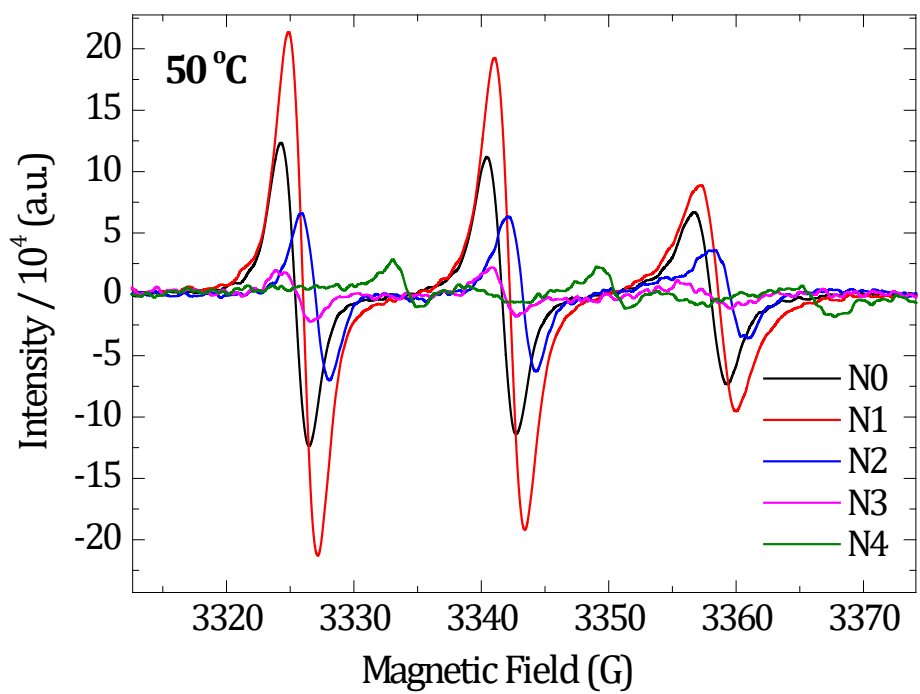

Figure S.4.- Set of ESR spectra for liposomes of DOPC:DODAB (70:30) coated with different number of layers as were obtained at two different temperatures (25º C and 50 ºC).